\documentclass[twocolumn]{aastex631}

\usepackage{amssymb}
\usepackage{amsmath}
\usepackage[caption=false]{subfig}
\graphicspath{{figures/}}

\newcommand{\brunt}{Brunt-V\"ais\"al\"a }
\newcommand{\mearth}{M$_{\oplus}$}
\newcommand{\Zsolar}{Z$_{\odot}$}

\begin{document}

%\title{Exploring Gas Giants: Concurrent Analysis of Jupiter and Saturn Evolution}
\title{Simultaneous Evolutionary Fits for Jupiter and Saturn Incorporating Fuzzy Cores}

\correspondingauthor{Ankan Sur}
\email{ankan.sur@princeton.edu}

\shorttitle{Simultaneous Fits to Jupiter and Saturn}
\shortauthors{Sur et al.}

\author[0000-0001-6635-5080]{Ankan Sur}
\author[0000-0001-6708-3427]{Roberto Tejada Arevalo}
\author[0000-0001-8283-3425]{Yubo Su}
\author[0000-0002-3099-5024]{Adam Burrows}
\affiliation{Department of Astrophysical Sciences, Princeton University, 4 Ivy Lane,
Princeton, NJ 08544, USA}

\begin{abstract}
With the recent realization that there likely are stably-stratified regions in the interiors of both Jupiter and Saturn, we construct new non-adiabatic, inhomogeneous evolutionary models with the same microphysics for each that result at the present time in respectable fits for all major bulk observables for both planets. These include the effective temperature, radius, atmospheric heavy-element and helium abundances (including helium rain), and the lower-order gravity moments $J_2$ and $J_4$. The models preserve from birth most of an extended  ``fuzzy" heavy-element core. Our predicted atmospheric helium mass fraction for Saturn is $\sim$0.2, close to some measured estimates, but in disagreement with some published predictions. To preserve a fuzzy core from birth, the interiors of both planets must start out at lower entropies than would be used for traditional ``hot start" adiabatic models, though the initial exterior mantle entropies can range from hot to warm start values. We do not see a helium ocean in Saturn's interior, and both models have inner envelopes with significant \brunt frequencies; this region for Saturn at the current epoch is more extended and in it, the Brunt is larger. The total heavy-element mass fraction in Jupiter and in Saturn is determined to be $\sim$14\% and $\sim$26\%, respectively, though there is some play in these determinations.   
\end{abstract}
    
\section{Introduction}

Early evolutionary models for Jupiter and Saturn assumed homogeneous and adiabatic interiors due to efficient convection \citep{Hubbard1968, Hubbard1969, Hubbard1970, Burrows1997, Baraffe1998, Burrows2001, Baraffe2003}. While this assumption successfully accounted for Jupiter's observed luminosity, it did not adequately reproduce Saturn's effective temperature at the current epoch. \citet{Stevenson1977a} addressed this discrepancy by proposing that hydrogen and helium in Saturn's interior were immiscible, leading to helium rainout. This rainout mechanism converts gravitational energy into internal energy and (perhaps) provides a latent heat, slowing Saturn's outer cooling rate. Further evidence supporting this theory emerged from the \textit{Galileo} entry probe, which measured helium depletion in Jupiter's atmosphere at a mass fraction value of $0.234\pm0.005$ \citep{vonZahn1998} compared to the protosolar value of 0.2777 \citep{Bahcall2006,Serenelli2010}. Consequently, this led to the development of non-homogeneous models, which incorporated helium gradients in the interiors of both Jupiter and Saturn and simultaneously provided respectable fits to their current effective temperatures \citep{Fortney2003, Mankovich2016, Pustow2016, Mankovich2020, Howard2024}. While the atmospheric helium abundance in Saturn remains uncertain due to the lack of direct measurement, it is estimated to range between 0.07 and 0.22 \citep{Conrath2000, Koskinen2018, Achterberg2020}.

Jupiter and Saturn were henceforth modeled as consisting of three distinct layers: A compact core with mass $M_c$, an outer convective envelope characterized by a helium mass fraction $Y_1$ and heavy element content $Z_1$, and an inner convective envelope defined by $Y_2$ and $Z_2$ \citep{Fortney2003, Saumon2004, Nettelmann2012, Nettelmann2013, Helled2013, Pustow2016, Movshovitz2020}. However, these updated models themselves have recently been challenged following the launches of NASA's \textit{Juno} \citep{Bolton2017a} and \textit{Cassini} \citep{Spilker2019} spacecraft, which among other things provided detailed gravitational moments for each planet \citep{Iess2018, Iess2019,Durante2020}. These observations suggest that Jupiter in particular may possess an extended region in its deep interior enriched with heavy elements, often referred to as a dilute or fuzzy core \citep{Wahl2017, Debras2019, Nettelmann2021, Militzer2022, Militzer2024}. \textit{Cassini} studies of Saturn's C ring have also strongly suggested that Saturn is undergoing gravity-mode pulsations, indicating that a substantial portion of Saturn's interior is stably-stratified by composition gradients and also possesses a fuzzy core of some sort in its interior \citep{Leconte2012, Mankovich2021, Nettelmann2021, Dewberry2021}. In this regard, \cite{Mankovich2021} have calculated a Saturn \brunt frequency profile, indicating where Saturn may be non-convective. 

\citet{Vazan2015} and \citet{Vazan2016} were the first to conduct evolutionary models of giant planets with convective heat transport and varying composition gradients, showing that the initial internal structure significantly affects the planet's subsequent evolution. More recently, \citet{Knierim2024} performed a similar study, exploring how primordial entropy profiles affect convective mixing and atmospheric composition in planets ranging from one Saturn mass to two Jupiter masses. Their results indicate that low primordial entropies and deep heavy-element mass fraction profiles can result in stable interior heavy-element gradients over evolutionary timescales. While these models provide valuable insights, they do not fit the structures of Jupiter or Saturn.

Currently, no evolutionary models exist that have simultaneously reproduced Jupiter and Saturn’s effective temperature, atmospheric helium abundance, and radius at 4.56 gigayear (Gyr) using fuzzy core models informed by \textit{Juno} and \textit{Cassini} data. \citet{Mankovich2020} are the first to fit Jupiter and Saturn simultaneously using a unified model with helium rain. However, these models do not account for fuzzy cores or match the gravity data and assume adiabaticity. \citet{Vazan2018} were the first to model Jupiter’s evolution with a primordial structure featuring a steep interior heavy-element gradient with a total heavy-element mass of 40 $M_{\oplus}$. While their model provided good fits for Jupiter’s effective temperature, radius, and gravitational moment $J_2$, it did not include helium rain and, therefore, could not match the observed atmospheric helium abundance. \citet{Muller2020} modeled Jupiter's evolution using initial conditions based on formation models, but their results failed to reproduce the dilute core inferred from \textit{Juno} data, while also matching Jupiter's effective temperature and radius. The most detailed study to date, which successfully matches Jupiter’s current effective temperature, radius, and helium fraction, but which also has a surviving fuzzy core, is by \citet{TejadaArevalo2024b}. That work used the latest equation of state (EOS) for hydrogen and helium mixtures \citep{Chabrier2021}, the latest atmospheric boundary conditions \citep{Chen2023}, and a full treatment of helium miscibility.

In this work, we aim to build on the work of \citet{Tejada2024} and fit Jupiter and Saturn simultaneously using models with surviving fuzzy cores that are also consistent to a reasonable level of precision with both \textit{Juno} and \textit{Cassini} gravity data and the inferred presence from C-ring ``Kronoseismology" of a partially stably-stratified interior in Saturn. Our approach incorporates the same helium rain prescription  and hydrogen-helium miscibility data \citep{Sur2024_apple}, reasonable total heavy element masses, the \cite{Chabrier2021} EOS, and the latest atmospheric boundary conditions \citep{Chen2023}. Starting from reasonable initial conditions, we seek to match the current effective temperatures, elemental abundances ($Y$ and $Z$), equatorial radii, and the gravitational moments ($J_2$ and $J_4$) of both planets, while also preserving an extended fuzzy core structure to the present epoch.

This paper is organized as follows: Section \ref{se2c:method} introduces the \texttt{APPLE} code. Section \ref{sec3:comparisons} demonstrates that \texttt{APPLE} can reproduce the previous results for both Jupiter and Saturn without fuzzy cores of \citet{Mankovich2020} and of \citet{Howard2024} using their EOSes and physical assumptions. Then, in Section \ref{sec4:models} we present our best-fit models for Jupiter and Saturn, including fuzzy cores and the latest physical inputs and ideas. Section \ref{sec4:conlusions} concludes with a discussion of the salient results that emerge from our new simultaneous model fit to both planets and outlines directions for future research and improvements.

\section{\texttt{APPLE} Code} 
\label{se2c:method}

We use \texttt{APPLE} \citep{Sur2024_apple} for computing all evolutionary models. Employing an operator-split method, it solves the hydrostatic structure with energy and species transport implicitly in time, governed by the equations:

\begin{align}
\frac{dP}{d M_r} &= -\frac{G M_r}{4\pi r^4} + \frac{\Omega^2}{6\pi r}
\label{eq:1}\\
\frac{dr}{d M_r} &= \frac{1}{4\pi r^2 \rho}
\label{eq:2}\\
\frac{\partial L}{\partial M_r} &= -T\frac{dS}{dt} - \sum_i \left(\frac{\partial U}{\partial X_i}\right)_{s,\rho}\frac{dX_i}{dt}
\label{eq:3}\\
N_A\frac{dX_i}{dt} &= -\frac{\partial }{\partial M_r}\left(4\pi r^2 F_i\right)\, ,
\label{eq:4}
\end{align}
where $X_i=\{X, Y, Z\}$ denotes the hydrogen, helium, and heavy element mass fractions, respectively, and all other symbols have standard definitions \cite[see][]{Sur2024_apple}. {Heavy elements in the envelope were modeled using the AQUA EOS \citep{Haldemann2020}, while a compact core of mass $M_c$, is made up of 50\% iron and
50\% post-perovskite (MgSiO$_3$) \citep{Keane1954, Stacey2004, Zhang2022}, is embedded in the center.} The code supports various H-He EOS options from the EOS module of \cite{Tejada2024}, such as SCvH95 \citep{Scvh1995}, CMS19+HG23 \citep{Chabrier2019, Howard2023}, and CD21 \citep{Chabrier2021}, with heavy elements mixed via the volume addition law. {The angular frequency in Equation \ref{eq:1} evolves over time, while conserving angular momentum. The time-dependent moment of inertia is derived from the structure using the Theory of Figures to Fourth Order \citep[TOF4;][]{Nettelmann2017}. The conservation calculation assumes present-day inferred rotation rates and corresponding moment of inertia estimates \citep[e.g.,][]{Militzer2022}.}

The miscibility module can incorporate curves from \citet{Lorenzen2009, Lorenzen2011}, \citet{Schottler2018}, and \citet{Brygoo2021} (hereafter referred to LHR0911, SR18, and Brygoo21, respectively), with adjustable temperature shifts and the full $Y$ dependence. Helium rain is simulated using \textit{scheme B} in \citet{Sur2024_apple}. For Jupiter, we use atmospheric boundary conditions from \citet{Chen2023}, and for Saturn we have developed new boundary conditions following the approach in \citet{Chen2023}. Energy transfer mechanisms include convection (via mixing-length theory), thermal conduction, and (quite subdominant, except in the atmosphere) radiation. Opacities for the atmosphere calculations are from \citet{Sharp2007} and \citet{Lacy2023}, and thermal conductivity is based on \citet{French2012}, adjusted to align with \citet{Zaghoo2017}.

\begin{figure*}
\centering
\includegraphics[scale=0.385]{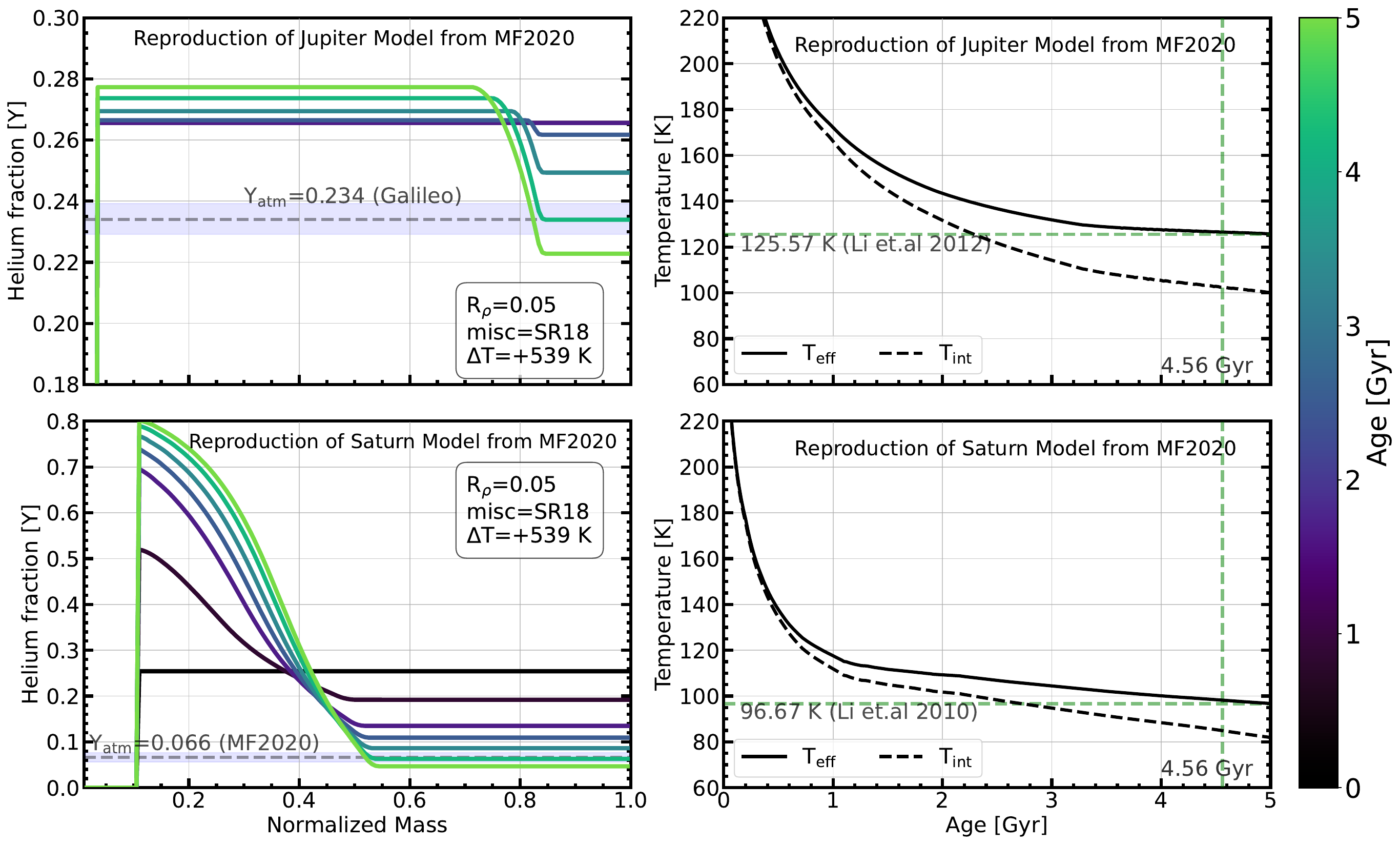}
\caption{Reproduction of the \citet[MF2020]{Mankovich2020} results for Jupiter and Saturn using \texttt{APPLE}. The left panels depict the helium abundance in the H-He envelope of the planet with the colorscale indicating the age in Gyr. The dashed line in the top left panel represents the \textit{Galileo} constraint $Y_{\rm atm}=0.238\times(1-Z_{\rm atm})=0.234$ \citep{vonZahn1998} while the dashed line in the bottom left panel shows the total helium mass fraction $Y_{\rm atm}=0.066$ corresponding to $Y_{\rm atm}/(1-Z)=0.07$, as obtained by MF2020. The right panels display the evolution of effective and internal temperatures with age, with the dashed line indicating the constraints from \cite{Li2010, Li2012} for both planets.}
\label{fig:comparison_MF2020}
\end{figure*}
\section{Reproducing recent results} 
\label{sec3:comparisons}

\subsection{Mankovich \& Fortney (2020)}
\label{mfort}

\cite{Mankovich2020} developed helium rain models for Jupiter and Saturn using an evolutionary code based on \cite{Thorngen2016}, assuming a protosolar helium mass fraction $Y_{\rm proto}=0.27$, and using the \citet[MH13]{Militzer2013} EOS for the H-He envelope. Heavy elements were assumed to be pure water ice modeled using the Rostock water EOS \citep{French2009}. The \cite{Fortney2011} boundary conditions were implemented, modifying that reference to assume a Bond albedo of 0.5. Their parametric study, incorporating the \cite{Schottler2018} miscibility curve, concluded that a temperature shift of $+539$ K and setting $R_{\rho}=0.05$\footnote{$R_{\rho}>0$ is referred to as superadiabatic, while $R_{\rho}$=0 is adiabatic, meaning the Schwarzschild condition is used to identify convective zones.} were necessary to simultaneously fit the effective temperatures of Jupiter and Saturn and the atmospheric helium abundance for Jupiter. \cite{Mankovich2020} also presented their helium fractions as $Y^{\prime}=Y/(1-Z)$ which is not the total helium mass fraction, but is the helium mass fraction relative to the total mass fraction of only hydrogen and helium. Using this definition, their results predicted that the atmospheric helium fraction for Saturn depleted to $Y^{\prime} = 0.07\pm 0.01$ at its current age.

In their models, \cite{Mankovich2020} varied the core masses ($M_c$) from 0 to 30 $M_{\oplus}$, and the uniformly-distributed heavy element mass fraction ($Z$) from 0 to 0.5, but did not specify their best-fit parameters for $M_c$ and $Z$. The \citet{Fortney2011} boundary conditions assume a $3 \times$ solar metallicity for Jupiter and $10\times$ solar metallicity for Saturn. However, it is unclear if these values were used in their hydrogen-helium envelope models. Therefore, in trying to reproduce their results using \texttt{APPLE}, we adopt $M_c=10 \, M_{\oplus}$ and an $M_z$ of $5 \, M_{\oplus}$ for both Jupiter and Saturn. This corresponds to $Z_{\rm env}=0.016$ for Jupiter and $Z_{\rm env}=0.058$ for Saturn, mixed uniformly with the envelope H-He. Their evolutionary models assumed an equilibrium temperature ($T_{\rm eq}^4$) to factor in insolation and used the relation $T_{\rm eff}^4 = T_{\rm int}^4  + T_{\rm eq}^4$ (again, with a Bond albedo of 0.5) to derive the effective emission temperature. $T_{\rm int}$ is the internal flux temperature that tracks the energy flux out of the core \citep{Sur2024_apple}.

\begin{figure*}
    \centering
    \includegraphics[width=0.98\linewidth]{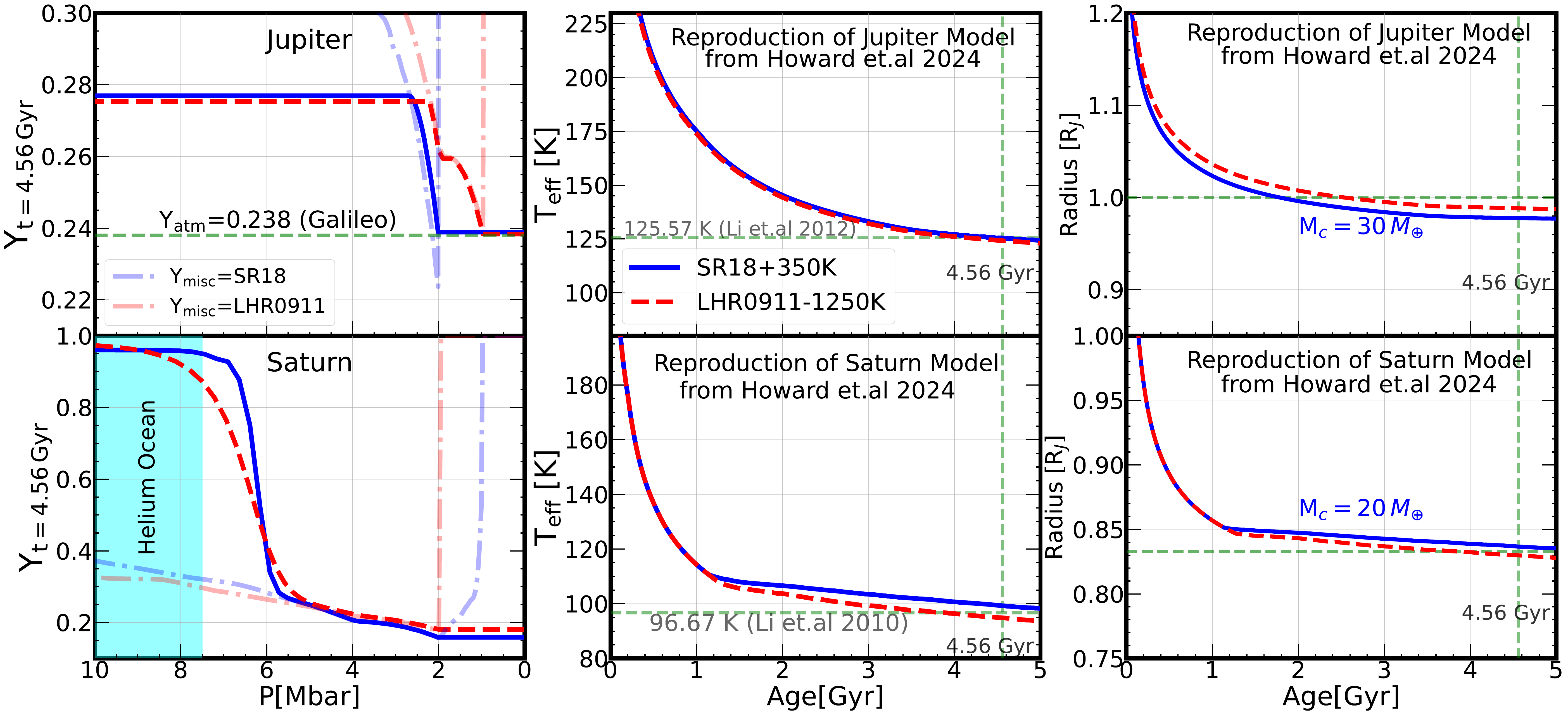}
    \caption{Reproduction of the \cite{Howard2024} results using \texttt{APPLE} with the SR18 (blue) and LHR0911 (red) miscibility curves, with their suggested temperature shifts (+350 K and -1250 K, respectively). Our comparison models, represented by solid lines, include a core of mass $30 \, M_{\oplus}$ for Jupiter and 20 $\,M_{\oplus}$ for Saturn. The CMS19+HG23 EOS is used for the H-He envelope and the models are evolved with the \cite{Fortney2011} boundary conditions (as did they). The left panels depict the helium mass fraction profile in the H-He envelope at 4.56 Gyr. We see the formation of a helium ocean at pressures above 7$-$8 Mbar. The dashed line shows the \textit{Galileo} $Y_{\rm atm}$ constraint \citep{vonZahn1998}. The middle panels depict the evolution of effective and internal temperatures with age, with the dashed line indicating the constraints on $T_{\rm eff}$ from \cite{Li2010,Li2012} for both planets. The right panels show the evolution of radius with age.}
    \label{fig:Howard2024}
\end{figure*}

Using \texttt{APPLE} \citep{Sur2024_apple}, we now attempt to reproduce the results of \citet{Mankovich2020}, using their setup (in particular their miscibility shift of +539 K), almost the same EOS, and their parameters. We employed the CD21 EOS for the H-He mixture, which at $Y^{\prime} = 0.245$ is the same as the MH13 EOS they used. They aimed to achieve at 4.56 Gyr an effective temperature of 125.57 K for Jupiter and of 96.67 K for Saturn \citep{Li2010, Li2012}. Using our helium rain algorithm, which differs slightly from their method of setting $Y$ in the helium rain region to the local equilibrium helium abundance, we obtained a total atmospheric helium mass fraction of $Y_{\rm atm}=0.233$ for Jupiter and $Y_{\rm atm}=0.061$ for Saturn\footnote{Note that this corresponds to a $Y^{\prime}$ of 0.237 for Jupiter and 0.066 for Saturn.}.  Moreover, at 4.56 Gyr and using an $\mathcal{H}_r=1\times10^8$ cm (where $\mathcal{H}_r$ sets the scale height of the helium rain region in our models), we obtain a $T_{\rm eff}$ of 126.53 K for Jupiter and (using $\mathcal{H}_r=2\times10^8$ cm) a $T_{\rm eff}$ of 98.2 K for Saturn. These numbers are consistent with the findings of \citet{Mankovich2020} under their model assumptions. The left panels in Figure \ref{fig:comparison_MF2020} illustrate the evolution of the total helium mass fractions at different times for both planets, while the right panel shows the corresponding evolution of $T_{\rm eff}$ and $T_{\rm int}$. As a reference, the dashed line in the bottom left panel shows $Y_{\rm atm}=0.066$, which corresponds to the $Y_{\rm atm}^{\prime} = Y_{\rm atm}/(1-Z)=0.07$ in \citet{Mankovich2020}.
We see in Figure \ref{fig:comparison_MF2020} that we reproduce the \citet{Mankovich2020} model quite well using \texttt{APPLE} and their physical inputs and assumptions.

\subsection{Howard, M$\ddot{u}$ller \& Helled (2024)}
\label{hmh24}

\cite{Howard2024} investigated the effects of various miscibility curves $—$LHR0911, SR18, and Brygoo21 $—$ on the adiabatic evolution of Jupiter and Saturn, including helium rain and $R_{\rho}=0$. Their baseline models assume all heavy elements are concentrated in the core, with core masses of $30 \, M_{\oplus}$ for Jupiter and $20 \, M_{\oplus}$ for Saturn. In the envelope, the CMS19+HG23 EOS was used for a pure H-He mixture. The flux boundary conditions from \cite{Fortney2011} were applied. To match Jupiter's atmospheric helium abundance of $Y_{\rm atm}^{\prime}=0.238$ at 4.56 Gyr, they found miscibility curve temperature offsets of -1250 K for LHR0911, +350 K for SR18, and -3850 K for Brygoo21 were necessary. With these adjustments, their Saturn models produced an atmospheric helium mass fraction of 0.13-0.16. Based on their miscibility curves, their Jupiter models yielded a $T_{\rm eff}$ $\sim$1.5 K lower than the current estimate, while the Saturn models differed by $\sim$2 K. 

We reproduce the results of \cite{Howard2024} using \texttt{APPLE} with their parameters, as shown in Figure \ref{fig:Howard2024}. However, our models differ in several minor ways. First, we included the chemical potential term in the energy equation, which is typically absent in the CEPAM evolution code used in their study \citep{Guillot_1995_CEPAM}. Second, our helium rain algorithm\footnote{The different helium rain schemes are described in the \texttt{APPLE} code paper \citep{Sur2024_apple}.} differs by employing a generalized diffusion equation in the flux-conservative form for helium redistribution, rather than setting it ``by hand" in the helium rain region to the helium abundance on the miscibility curve at the local temperatures and pressures. To anticipate the formation of the helium ocean, as observed by \citet{Howard2024}, we made a slight adjustment to our helium rain \textit{scheme B} in \texttt{APPLE}. This minimal modification allows the natural formation of a helium ocean by setting a (high) maximum threshold value for $Y$ in any mass zone; when the helium mass fraction (set here to 0.95) exceeds this value and the region thereby becomes supersaturated, helium is naturally redistributed by the diffusion algorithm into the zones above. Lastly, to align our evolutionary models with those of \citet{Howard2024}, we disabled core heat transport. Including core heat transport in these models results in a 0.4\% increase in $T_{\rm eff}$, a 2\% increase in $Y_{\rm atm}$, and a 0.6\% decrease in the radius.

% the species diffusion equation is now given by
% \begin{multline}
% \frac{dY}{dt} = \frac{\partial }{\partial M_r}\left\{4\pi r^2 \rho \Dmlt \frac{\partial Y}{\partial r}\right.\\
% \left.+ \frac{4\pi r^2 \rho \Dmlt}{\mathcal{H}_r} \big(Y - Y_{\rm misc}\big)_{+} \left(1 - \frac{Y}{Y_{\rm th}}\right)_{+}\right\} \, ,
% \label{eq:misc2}
% \end{multline}
% where $Y_{\rm th}$ is the maximum threshold in the helium ocean, $Y_{\rm misc}$ is the lower intercept of the planetary T-P profile with the miscibility curve, and the $+$ indicates only positive values of the expression are evaluated. A mass zone that becomes supersaturated deposits its helium in the zone above it. Diffusion coefficients are calculated using the Mixing-Length Theory and Schwarzschild criterion for convection. 

In Figure \ref{fig:Howard2024}, we plot (left) the helium mass fraction profile at 4.56 Gyr and the evolution of the $T_{\rm eff}$ (middle) and radius (right) of Jupiter (top row) and Saturn (bottom row), employing our variant of the setup and parameters of \citet{Howard2024}. We get a $T_{\rm eff}$ of 125.04 K and 125.03 K for shifted SR18 and LHR0911, respectively.  For Jupiter, we used our \textit{scheme B} for helium rain by setting $H_r\sim4\times10^7$ cm for SR18, and $H_r\sim5\times10^6$ cm for LHR0911, both producing $Y_{\rm atm}^{\prime}=0.238$ at 4.56 Gyr, as shown in the top-left panel of Figure \ref{fig:Howard2024}.  

To replicate the \citep{Howard2024} Saturn model, we used the modified \textit{scheme B} and set $\mathcal{H}_r=5\times10^7$ cm for LHR0911 and $\mathcal{H}_r=2\times10^7$ cm for SR18, yielding $Y_{\rm atm}^{\prime}=0.175$ and $Y_{\rm atm}^{\prime}=0.158$, respectively. These are consistent with what was found in the models of \citet{Howard2024}. Our version of their SR18 model for Saturn gives a $T_{\rm eff}$ of 99.2 K, while for the LHR0911 model, we derive a $T_{\rm eff}$ of 96.2 K. We observe the onset of helium rain at 3.64 Gyr for Jupiter and 1.3 Gyr for Saturn, which \citet{Howard2024} also found. We replicate the formation of a helium ocean above the core at pressures exceeding 8 Mbar, with the helium fraction in this region enriched to $Y_{\rm th}=0.98$ and spanning 20\%-30\% of Saturn's total mass. Lastly, we obtained a Jupiter radius of 69,115.41 km and a Saturn radius of 58,165.95 km, which are close to the mean radii of the individual planets reported by \cite{Howard2024}.

Hence, despite slight differences in numerical method and various codes, when employing the parameters and setups of previously published models for Jupiter and Saturn \citep[e.g.,][]{Mankovich2020,Howard2024}, we can with some facility capture all the general results of these models. We now proceed to provide our updated non-homogeneous and non-adiabatic evolutionary models for both planets in the context of the current presence of fuzzy cores and stably-stratified interiors inferred for both Jupiter and Saturn.

\begin{figure*}
    \hspace{-0.4cm}\includegraphics[width=1.05\linewidth]{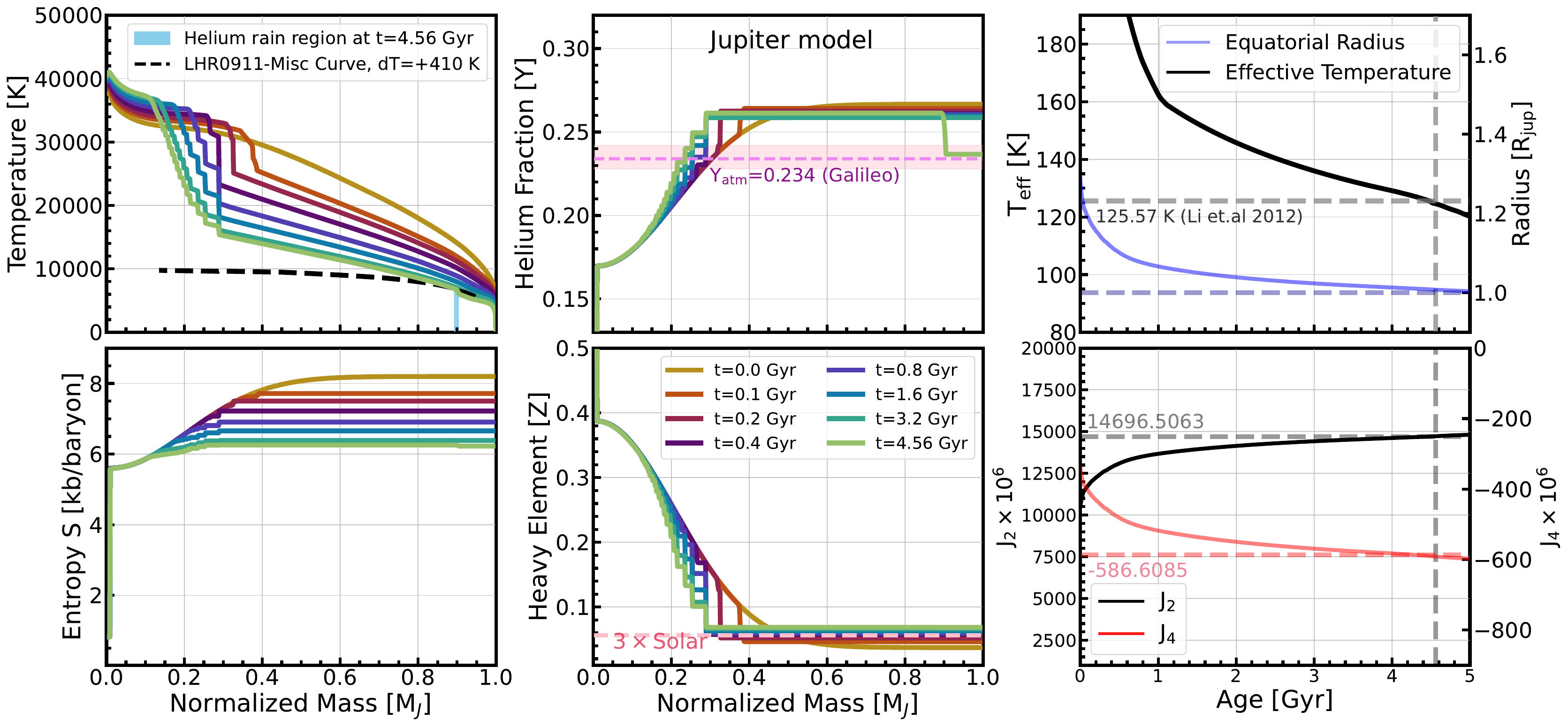}
    \caption{A very slightly improved evolutionary model for Jupiter  \cite[see][]{TejadaArevalo2024b} with an initial fuzzy core that matches within observational uncertainties the current values of the effective temperature, equatorial radius, atmospheric helium abundance, outer envelope metallicity, and $J_2$ and $J_4$ gravitational moments. The initial outer entropy is 8.2 k$_{\rm B}$/baryon, while the interior entropy is 7.5 k$_{\rm B}$/baryon. The panels display: (top left) the evolution of the temperature profile; (top middle) the evolution of the helium mass fraction ($Y$) profile; (top right) effective temperature and equatorial radius; (bottom left) evolution of the entropy profile; (bottom middle) evolution of the $Z$ profile; and (bottom right) evolution of the gravitational moments $J_2$ and $J_4$. This Jupiter model contains 42.5 \mearth of heavy elements, including a compact core of 3 \mearth ($Z = 1$). As shown in the bottom middle panel, the initial $Z$ profile becomes convectively mixed from the outer layers inward, with the most significant changes occurring early in the evolution. The outer metallicity starts at 1.9 \Zsolar\ and increases to 3.6 \Zsolar. Helium rain begins at 4 Gyrs, based on the LHR0911 miscibility curves with a +410 K temperature shift, resulting in outer helium depletion to $Y = 0.236$, consistent with \textit{\textit{Galileo}} entry probe measurements \citep{vonZahn1998}. The black dashed line in the top left panel indicates the demixing temperature curve encountered by the cooling temperature profile, and the small vertical blue-shaded region highlights the helium rain zone. The time stamps are indicated in different colors on the bottom middle panel.}
    \label{fig:jupiter_best_fit}
\end{figure*}

\section{New Jupiter and Saturn Evolutionary Models}
\label{sec4:models}

The results from the \textit{Juno} probe \citep{Bolton2017a, Bolton2017b, Wahl2017, Debras2019, Nettelmann2021, Miguel2022, Militzer2022, Howard2023, Militzer2024} imply that Jupiter has a large ``fuzzy" core that may be stably stratified, and, hence, non-convective and non-adiabatic. Moreover, the \textit{Cassini} probe \citep{Iess2019, Spilker2019} has measured various propagating modes in Saturn's C-ring. If the excitation of these modes is indeed caused by gravitational coupling with modal pulsations of Saturn (otherwise not yet directly seen), since some of these modes have frequencies that suggest the presence in Saturn of g-modes and g-modes are evanescent in convective zones, this indicates the existence of a large stably-stratified region in the interior of Saturn as well. 

The upshot is that, despite years of modeling both planets under the assumption that they are adiabatic and fully convective, the \textit{Juno} and \textit{Cassini} data are now strongly indicating that this is not true. Therefore, modern models of the evolution of giant planets must incorporate, in a way not done in the past, both stably stratified and convective regions. This is necessary to comport with all the measured physical constraints for both Jupiter and Saturn and is one major motivation for the current paper. In addition, the atmosphere boundary conditions for both planets have recently been updated to include ammonia clouds, to naturally reproduce the updated measurements of their Bond albedos, and to be more consistent with the effects of solar insolation \citep{Chen2023}. We now discuss our exploration of such models using our code \texttt{APPLE}.

\subsection{Jupiter Model}
\label{jupiter}

In our recent work \citep{TejadaArevalo2024b}, we developed the first evolutionary model for Jupiter that incorporates helium rain and a fuzzy core of heavy elements surviving over 4.56 Gyr. This model successfully reproduces Jupiter's observed effective temperature of $\sim 125$ K \citep{Li2012}, an equatorial radius of $\sim$71,697 km, atmospheric helium abundance of 0.234 \citep{vonZahn1998}, and atmospheric metallicity of $\sim$3$\times$solar (\Zsolar). We explored various initial distributions of heavy elements, including flat-top and Gaussian profiles with different radial extents, as well as the impact of ``hot" and ``cold" start entropy profiles. Key findings from that study were:
\begin{itemize}
    \item In the outer envelope, the composition gradients are generically flattened by convection, leading to mixing and homogenization. 
    \item A lowish initial interior entropy near $\sim$${S=7.5}$ k$_{\rm B}$/baryon is necessary to sustain a fuzzy core. The initial surface entropy, set at 8.2 k$_{\rm B}$/baryon, has minimal impact on long-term evolution, as the influence of the outer initial entropy dissipates within a few million years.
    \item In part to match Jupiter's measured equatorial radius, the model contains $\sim$14 \mearth\ of heavy elements in the core and 30 \mearth\ in the envelope, totaling 44 \mearth.
    \item The fuzzy core extends to 42\% of the planet's radius ($\sim 0.4 M_J$) at 4.56 Gyr. A core that extends beyond this would either mix or lead to effective temperatures colder than 125 K.
    \item A shift of +300 K for the LHR0911 miscibility curve was required to match Jupiter's observed atmospheric helium abundance. 
\end{itemize}
  
Building on these results, we present an updated Jupiter model that retains the features of \cite{TejadaArevalo2024b}, while also now {\bf approximately} reproducing the gravitational moments $J_2$ and $J_4$ \citep{Durante2020}. This updated model benefits from enhanced numerical smoothing of thermodynamic derivatives in our equation of state module \citep{Tejada2024}. Gravitational moments are calculated using the Theory of Figures to fourth order \citep[ToF4,][]{Nettelmann2017}, which also provides the moment of inertia and rotation-corrected equatorial radius. These quantities are important in the centrifugal term of the equation of hydrostatic equilibrium (eq. \ref{eq:1}), under the assumption of solid-body rotation, to provide a second-order correction to account for model oblateness. {We do not normalize either to the present-day (4.56 Gyrs.) equatorial radius, but let the code arrive naturally at values of radius, $J_2$, and $J_4$.}

We conducted a grid search of more than 2,000 models using \texttt{APPLE} to identify parameter combinations that produce a Jupiter-like planet, subject to observational constraints. The core mass was varied from 1 to 15 \mearth, the total heavy element mass from 36 to 44 \mearth, and the LHR0911 miscibility curve was shifted between +300 K and +450 K in steps of 10 K. The standard deviation of a Gaussian profile was adjusted to vary the extent of the heavy element distribution. The entropy at the outer boundary was fixed at 8.2 k$_{\rm B}$/baryon, and the initial helium fraction $Y^{\prime}=0.277$ was set uniformly in the envelope. The models employed the CD21 equation of state (EOS) for H-He \citep{Chabrier2021}, the LHR0911 miscibility curve for helium rain (plus variable temperature shifts), and the \cite{Chen2023} atmospheric boundary conditions. A key feature of \texttt{APPLE} is that the miscibility curve adjusts itself as $Y^{\prime}$ changes and is not held fixed during the evolution. We use the Ledoux criterion for convective stability in our heat transport and set Jupiter's current rotation period at 9:55:29 (hr:min:s) \citep{Yu2009}. Our \texttt{APPLE} simulations used 750 mass zones with a timestep tolerance of 1\%.

\begin{figure*}
    \hspace{-0.4cm}\includegraphics[width=1.05\linewidth]{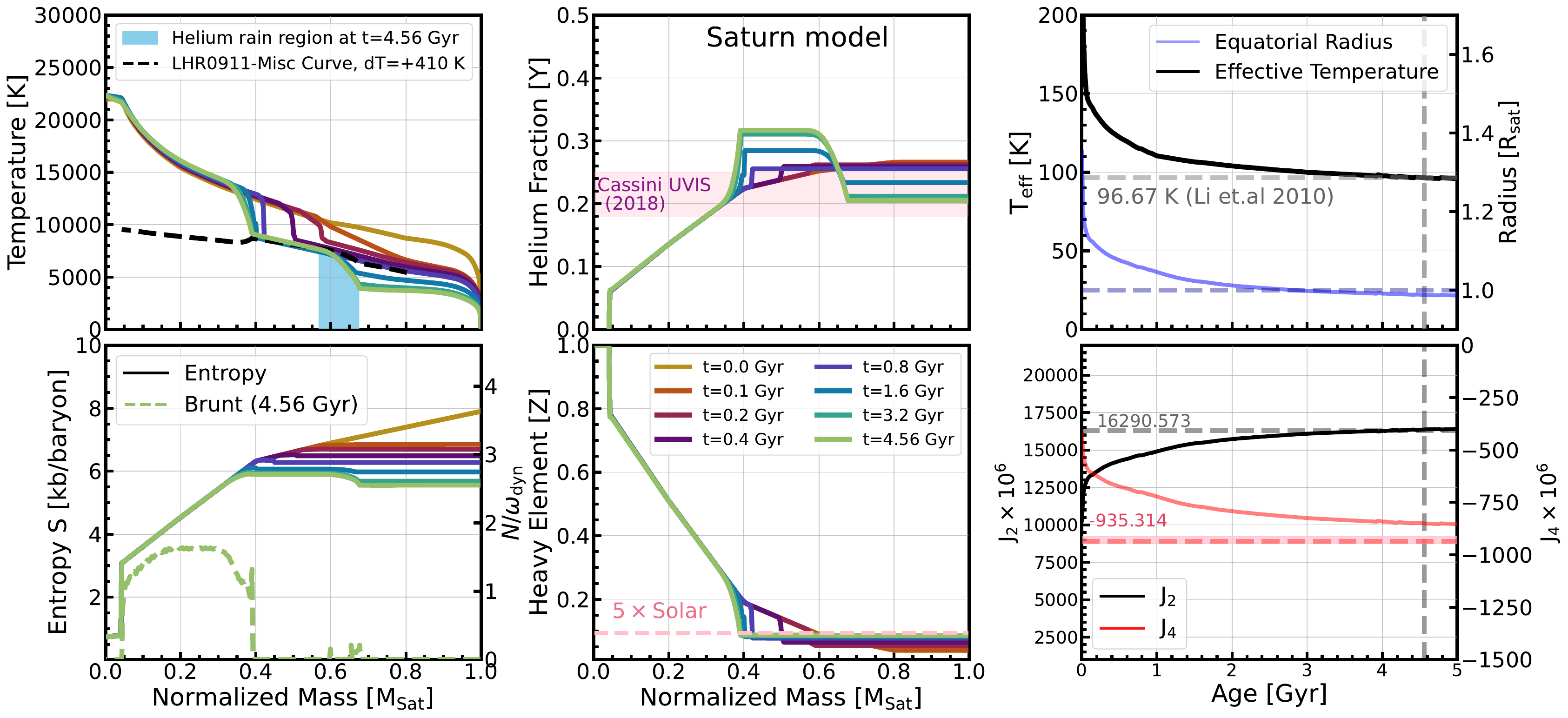}
    \caption{The best fit evolutionary model for Saturn with an initial fuzzy core that matches its current values of the effective temperature, equatorial radius, atmospheric helium abundance, outer envelope metallicity, \brunt frequency, and gravitational moments within observational uncertainties. The initial outer entropy is 7.9 k$_{\rm B}$/baryon, while the interior entropy $\sim 6.2$ k$_{\rm B}$/baryon. The panels are the same as in Figure \ref{fig:jupiter_best_fit}: (top left) evolution of the temperature profile; (top middle) evolution of the helium mass fraction ($Y$) profile; (top right) evolution of the effective temperature and equatorial radius; (bottom left) evolution of the entropy profile; (bottom middle) evolution of the $Z$ profile; and (bottom right) evolution of the gravitational moments $J_2$ and $J_4$. This model contains 25 \mearth\ of heavy elements, including a compact core of 4 \mearth.  The outer metallicity starts at 2.0 \Zsolar\ and increases to 4.57 \Zsolar. Helium rain begins at 0.92 Gyr, based on the LHR0911 miscibility curve with a +410 K temperature shift that also fits our Jupiter model, resulting in outer helium depletion to $Y_{\rm atm} \sim 0.205$, consistent with \cite{Conrath2000, Koskinen2018} (shown by the pink shaded region in top middle panel). The top left panel shows the miscibility curve and helium rain region. The \brunt ratio of approximately 1.7 at 4.56 Gyr, closely aligning with the value of 2 predicted by \cite{Mankovich2021}, is shown in the bottom left panel. Similar to \cite{Mankovich2021}, whose stably stratified region extends to 0.6 Saturn radii, in order to match Saturn's current observables best our model finds that the stable region extends to 0.5 R$_{\rm Sat}$.}
    \label{fig:saturn_best_fit}
\end{figure*}   
Figure \ref{fig:jupiter_best_fit} illustrates the evolution of our best-fit Jupiter model. It is very similar to that found by \cite{TejadaArevalo2024b}, with slight differences due to the slightly smoother thermal derivatives employed and the desire here simultaneously to fit the lower gravitational moments. This model predicts a total heavy element mass of 42.5 \mearth, of which 3 \mearth\ resides in the core with $Z=1$. To match Jupiter's observed $Y_{\rm atm}$, the required temperature shift for the LHR0911 miscibility curve is +410 K \citep[compared to +300 K in ][]{TejadaArevalo2024b} with helium rain initiating at 4 Gyr. The resulting best-fit model numbers with the corresponding percent differences (in parentheses) from the measured values are: current effective temperature $T_{\rm eff} = 124.6$ K ($\sim 0.7\%$), $Y_{\rm atm}=0.236$ ($\sim 0.8\%$), $Z_{\rm atm} = 3.6$ \Zsolar, and equatorial radius $=$ $R_{\rm eq} = 72019.5$ km ($\sim 0.7\%$). The gravitational moments $J_2$ and $J_4$ are $14731.6 \times 10^{-6}$ ($\sim 0.2\%$) and $-591.46\times 10^{-6}$ {($\mathbf{\sim 0.8\%}$)}, respectively. {Table \ref{tab:jupiter_saturn_comparison} presents a comparison of our results with the observed measurements for Jupiter.}

\subsection{Saturn Model}
\label{saturn}

\cite{Mankovich2021} explored Saturn's internal structure, focusing on its core, using seismic data from the planet's rings. Their ring seismology analysis revealed gravity mode (g-mode) pulsations, indicating stable stratification due to composition gradients in a large portion of Saturn's interior \citep[see also][]{Fuller2014}. The study identified a diffuse core-envelope transition region extending to about 60\% of Saturn's radius, containing approximately 15.5-20.8 \mearth of ice and rock. This tapered distribution of heavy elements points to a complex internal structure, offering insights into Saturn's primordial composition and accretion history. One of the main findings of that study was that in the stably-stratified region in Saturn the ratio of the \brunt frequency to the dynamical frequency, i.e. $N/\omega_{\rm dyn}$, is approximately 2.

Saturn's atmospheric helium abundance remains poorly known, with estimates agreeing on depletion relative to the protosolar value, but differing substantially in magnitude. These estimates range from low values of $Y_{\rm atm}$ = $0.02-0.13$ \citep{Conrath1984, Achterberg2020} to higher values of $Y_{\rm atm}$ = $0.18–0.25$ \citep{Koskinen2018,Conrath2000,Orton1980}. Such a span leaves the helium abundance of Saturn's atmosphere currently all but unconstrained.     

\begin{table*}
\centering
\caption{Comparison of measured quantities and our model results for Jupiter and Saturn at 4.56 Gyr. References for the measured values: [1] \citet{Li2010}, [2] \citet{Li2012}, [3] \citet{Seidelmann2007}, [4] \citet{vonZahn1998}, [5] \citet{Achterberg2020, Koskinen2018}, [6] \citet{Guillot2023}, [7] \citet{Iess2018}, [8] \citet{Iess2019}}
\label{tab:jupiter_saturn_comparison}
\begin{tabular}{lcccc}
\hline
\hline
Quantity & \multicolumn{2}{c}{Jupiter} & \multicolumn{2}{c}{Saturn} \\
         & Measured & Our Model (4.56 Gyr) & Measured & Our Model (4.56 Gyr)\\
\hline
T$_{\rm eff}$ [K]              & 125.57 $\pm$ 0.07 [1]  & 124.6  & 96.67 [2] & 96.54 \\
Equatorial Radius [km]          & 71492 $\pm4$ [3]  & 72019.5  & 60268 $\pm4$ [3] & 59,551.8  \\
Atmospheric Helium Fraction (Y$_{\rm atm}$)    & 0.234 $\pm0.005$ [4] & 0.236  & 0.075-0.22 [5] & 0.205 \\
Atmospheric Metallicity (Z/Z$_\odot$)          & 1.5--5 [6]   & 3.6    & 5.0--10.0 [6]  & 4.6 \\
$J_2$ $\times 10^{-6}$         & 14696.572 [7]  & 14731.6    & 16290.573 [8] & 16365.7 \\
$J_4$ $\times 10^{-6}$           & -586.609 [7]  & -591.46  & -935.314 [8]  & -850.11  \\
\hline
\end{tabular}
\end{table*}

We have constructed an initial Saturn model featuring a deep heavy-element gradient in the interior, mimicking a fuzzy core, with an entropy of $S=7.9$ k$_{\rm B}$/baryon at the surface and $S\sim6.2$ k$_{\rm B}$/baryon in the interior. This configuration establishes an initial value of $N/\omega_{\rm dyn} \sim 2$ in the diffuse core. Similar to the Jupiter model, we set the envelope helium/hydrogen ratio $Y^{\prime}$ equal to the proto-solar value $0.277$, and adopt the Jupiter-informed LHR0911 miscibility curve with a temperature shift of +410 K to model helium rain. The H-He mixture is modeled using the CD21 equation of state (EOS), and we apply the atmosphere boundary condition methodology of \citet{Chen2023}, under the assumption that the metallicity of Saturn's atmosphere is $5\times$solar. The Ledoux criterion for convective stability is used for heat transport  \citep{Sur2024_apple,Tejada2024}. We run a grid of 2,000 models for Saturn, employing a spatial resolution of 500 zones and a timestep tolerance of 1\%. Parameters varied include the radial extent of the heavy-element distribution from 0.4 to 0.6 times Saturn's initial radius, a total heavy-element mass of $Z_{\rm mass} \in [24,28]$ \mearth, core masses from 1$-$5 \mearth, and surface entropies of 7.4 to 8.0 k$_{\rm B}$/baryon\footnote{Note that in our current approach varying the surface entropy also shifts the interior entropies by the same amount.}. The current rotation period for Saturn is set to 10:33:34 \citep{Militzer2019, Militzer2023}. 

Figure \ref{fig:saturn_best_fit} illustrates the evolution of our best-fit Saturn model, which predicts a total heavy-element mass of 25 \mearth, with 4 \mearth\ located in the core. The model closely reproduces the observed values, yielding an effective temperature of 96.54 K, a deviation of 0.13\% from current measurements \citep{Li2010}, an atmospheric helium fraction of $Y_{\rm atm} = 0.205$, and an atmospheric metallicity of $Z_{\rm atm} \sim 4.57$ \Zsolar\ after 4.56 Gyr of evolution. At the current epoch, the equatorial radius is $\sim$59,551.8 km, deviating by roughly 1.18\%, while the gravitational moments are $J_2 = 16365.7 \times 10^{-6}$ and $J_4 = -850.11 \times 10^{-6}$, differing by 0.4\% and a small margin, respectively. This model successfully matches Saturn's observed properties, while offering insights into the sensitivity of its interior structure to variations in key parameters.  {Table \ref{tab:jupiter_saturn_comparison} provides a summary of our results alongside the observed measurements for Saturn.}

The following key findings in our model are:    
\begin{itemize}
    \item If a fuzzy heavy-element core survives from birth, then there is no helium ocean formed.
    \item Helium piles up not in the interior, but in an intermediate region bounded from above by the helium rain region.
    \item The predicted atmospheric helium abundance is $\sim$0.205, close to the measurements of \cite{Conrath2000,Koskinen2018}.  This is $\sim$3 times the prediction of \cite{Mankovich2020} and is in part a consequence of the possible existence of an inner stable fuzzy core at the current epoch that has survived since formation.
    \item We simultaneously achieve a current \brunt ratio of $\sim$2 in the interior $\sim$50\% of Saturn's radius, which in that region is stably stratified.
    \item The inner $\sim$40\% by mass, as it is not convective, barely cools on solar-system timescales. 
\end{itemize}

\section{Discussion and Conclusions}
\label{sec4:conlusions}

In this paper, we have attempted, using the best available EOSes, atmosphere cooling boundary conditions, hydrogen-helium miscibility physics, and our new \texttt{APPLE} planet evolution code \citep{Sur2024_apple} to create evolutionary models for both Jupiter and Saturn that simultaneously fit their observables, while preserving from birth a ``fuzzy" extended heavy-element core. Such a core is now strongly suggested for both planets and requires evolutionary models that do not assume adiabaticity nor chemical homogeneity. We find that the predicted mass fraction of helium in Saturn's atmosphere should be near 0.2, close to the measurements of \citep{Conrath2000, Koskinen2018}, but in contrast with the predictions of \citet{Mankovich2020}. The presence of a stable inner fuzzy core is the major reason for the higher atmospheric $Y$, since such a stable region is a barrier to outer envelope helium depletion and inward settling. We note, however, that $Y_{\rm atm}$ is not currently well-measured, so its definitive future determination would clearly help to verify or refute our conclusions.

Though we have attempted to use best practices and the latest microphysics \citep[e.g.,][]{Chen2023,Tejada2024,Sur2024_apple}, there remain many uncertainties to resolve. Foremost among them concern the equations of state and the miscibility physics. We have been forced by the absence of suitable published equations of state for mixtures to employ the ``volume addition law" to combine our hydrogen/helium equation of state with our metal equation of state \citep{Tejada2024} and we have assumed that the mantle ``metal" is water. At the future levels of model fidelity to which the field is likely moving, a more informed envelope metal composition would certainly be desired. Moreover, there remains a gulf between measured \citep{Brygoo2021} and theoretical \citep{Lorenzen2011,Schottler2018} hydrogen-helium miscibility curves. In addition, the latent heat for hydrogen-helium phase separation has not been published, and this may have a bearing on all aspects of helium rain physics \citep{Markham2024}. 

Furthermore, we have assumed that core rotation is solid-body, that angular momentum is conserved, and that the ToF4 formalism \citep{Nettelmann2017} adequately (for evolutionary purposes) captures the oblateness and the gravity moments. Though for Jupiter we are able to reproduce both $J_2$ and $J_4$ (along with all the other major observables), for Saturn we are slightly off in $J_4$, while still reproducing all its other major observables in the context of a non-adiabatic and compositionally inhomogeneous model. We note that we are not trying to achieve perfection in the latter, since evolving exactly to the current well-measured gravity moments from an arbitrary initial state would be more than amazing. Nevertheless, improvements in all model components are always desirable, when and if possible. We do emphasize that the precision with which the current gravity moments can be inverted to determine current mass density and rotation profiles using the \textit{Juno} and \textit{Cassini} gravity data does not translate into correspondingly precise constraints on the thermal interiors; the gravity data are all but mute on this important aspect of planetary structure.      

We emphasize that we have for this work used the Ledoux condition for convective stability and have ignored the possibility of semi-convection \citep{Chabrier2007,Rosenblum2011, Mirouh2012, Leconte2012, Wood2013, Medrano2014, Tulekeyev2024}. We feel that the physics of doubly-diffusive mixing and transport is still in flux and we will leave to future work our assessment of this important topic. It may well be that the Schwarzschild condition would be better, given convective overshoot \citep{Korre2019, Anders2022} and new insights into the devolution in some parameter regimes on short timescales of semi-convection into convection \citep{Garaud2017,Tulekeyev2024}. Moreover, the erosion of the core over time \citep{wilson2012a,wilson2012b,Moll2017,Oberg2019}, and not its production at formation, may be responsible for the current presence of a fuzzy core.  Finally, there is the intriguing possibility that rapid rotation in Saturn may itself inhibit the penetration of a convective zone into its interior, thereby potentially explaining (or facilitating) the survival of a fuzzy core \citep{Fuentes2023, Fuentes2024}. This possibility, along with the potential importance of a large miscibility latent heat \citep{Markham2024}, emphasizes that there is yet much to do to resolve the many important outstanding issues in Jovian planet structure and evolution. Nevertheless, we suggest that our new models, generated in the context of the new understanding of Jupiter and Saturn that has emerged in the last few years, are important first steps towards the ultimate revelation of their internal characters.

\section{Acknowledgment}   
%\begin{acknowledgments}
Funding for this research was provided by the Center for Matter at Atomic Pressures (CMAP), a National Science Foundation (NSF) Physics Frontier Center, under Award PHY-2020249. Any opinions, findings, conclusions, or recommendations expressed in this material are those of the author(s) and do not necessarily reflect those of the National Science Foundation. YS is supported by a Lyman Spitzer, Jr. Postdoctoral Fellowship at
Princeton University.
%\end{acknowledgments} 

\bibliography{references}{}
\bibliographystyle{aasjournal}

\end{document}